# Fiber-optic nonlinear wavelength converter for adaptive femtosecond biophotonics

Geng Wang[1,2], Jindou Shi[1,2], Rishyashring R. Iyer,[1,2] Janet E. Sorrells[1,3], Haohua Tu[1,2]*

[1]Beckman Institute for Advanced Science and Technology, University of Illinois at Urbana-Champaign, Urbana, IL 61801, USA

[2]Department of Electrical and Computer Engineering, University of Illinois at Urbana-Champaign, Urbana, IL 61801, USA

[3]Department of Bioengineering, University of Illinois at Urbana-Champaign, Urbana, IL 61801, USA

*Corresponding author. Email: htu@illinois.edu.

## Abstract

Broad and safe access to ultrafast laser technology has been hindered by the absence of optical fiber-delivered pulses with tunable central wavelength, pulse repetition rate, and pulse width in the picosecond-femtosecond regime. To address this long-standing obstacle, we developed a reliable accessory for femtosecond ytterbium fiber chirped pulse amplifiers, termed as fiber-optic nonlinear wavelength converter (FNWC), as an adaptive optical source for the emergent field of femtosecond biophotonics. This accessory embowers the fixed-wavelength laser to produce fiber delivered ~20 nJ pulses with central wavelength across 950-1150 nm, repetition rate across 1-10 MHz, and pulse width across 40-400 fs, with a long-term stability of >2000 hrs. As a prototypical label-free application in biology and medicine, we demonstrate the utility of FNWC in real-time intravital imaging synergistically integrated with modern machine learning and large-scale fluorescence lifetime imaging microscopy.

## Introduction

Ultrafast laser engineering has produced mode-locked optical pulses with (sub-)picosecond duration ($\tau$) at a MHz-level repetition rate ($f$)[1] and driven the broad field of femtosecond biophotonics[2]. These ultrashort pulses were first produced six years after the invention of laser[3]. Subsequently, the Ti:sapphire crystal was recognized as a better lasing medium than dye solutions to support a broad range of near-infrared wavelengths ($\lambda$)[4]. The revolutionary development of Kerr lens mode-locking in 1990[5] led to the commercialization of high average power ($P$) Ti:sapphire lasers tunable across 690-1020 nm. More recent innovation around 2010 resulted in an ytterbium-based optical parametric oscillator (OPO) with one output widely tunable across 680-1300 nm and another synergistic output fixed at ~1040 nm[6]. Today, automatic wavelength tuning, beam pointing correction, and dispersion compensation have enabled many aspects of femtosecond biophotonics, including biological microscopy or clinical imaging[7], nanosurgery[8], and optogenetics[9]. However, despite decades of development in solid-state ultrafast source, it remains technically challenging to tune the three pulse parameters of $\lambda$, $f$, and $\tau$ independently and widely with sufficient output $P$ or pulse energy ($E$). In particular, the typical inability to vary $f$ not only limits the solid-state ultrafast lasers themselves, but also the subsequent wavelength-tuning accessories of optical parametric amplifier (OPA).

In contrast to the solid-state lasers, ultrafast fiber lasers have played a relatively minor role in femtosecond biophotonics despite its rapid advances[10], largely due to the difficulty to tune $\lambda$ (and to a less degree, $\tau$ toward shorter durations). The fiber chirped pulse amplification (FCPA) [11] of a pulse-picked seed along a large-core ytterbium (Yb) gain fiber, which could be either a conventional circular fiber or a largely single-mode photonic crystal fiber such as DC-200/40-PZ-Yb (NKT photonics)[12], has led to various commercial FCPA lasers (Table S1) useful for LASIK eye surgery and material processing. These pulse-picked FCPA (pp-FCPA) lasers are advantageous over the solid-state lasers due to the ease to vary $f$ at the same $P$, i.e., pre-amplification pulse picking for variable $E$. It seems that pairing one pp-FCPA laser with an OPA accessory would empower the tuning of $\lambda$ and $\tau$ (Table S1) to compete favorably with the solid-state lasers. However, the OPA is a largely free-space add-on that diminishes the fiber-optic advantages of the pp-FCPA laser, e.g., high resistance to environmental disturbance and good beam quality ensured by single-mode fiber propagation. Also, routine operation and maintenance of an integrated FCPA-OPA laser is often beyond the expertise of a biologist. Thus, the tuning accessory based on OPA technology has limited the application of the otherwise attractive pp-FCPA lasers to compete with their solid-state counterparts. To overcome these OPA-related limitations, we aim to develop an alternative tuning accessory based on the seeding subunit of OPA technology[13], known as supercontinuum (or white-light) generation.

## Results

### *High peak-power coherent fiber supercontinuum generation*

Bulk-medium supercontinuum generation was demonstrated in glasses using ps pulses[14]. Later, fs pulses were more useful in the seed generation of commercial OPA operated at 0.25 MHz[15] (Table S2), which also enabled the commercial OPA accessories of the pp-FCPA lasers toward larger $f$ of ~4 MHz (Table S1). Interestingly, photonic crystal fiber-based supercontinuum generation was first demonstrated using fs pulses[16], but ps pulses gained commercial success later due to its robust all-fiber setup[17] (Fig. 1a, Approach 1; Table S2). The success of this ps approach in wide spectral broadening has largely restricted the fs approach to an add-on nonlinear wavelength converter for a solid-state Ti:sapphire oscillator (Fig. 1a, Approach 2). A third approach has diverged from either the all-fiber supercontinuum generation or the solid-state laser, and instead focused on coherent fiber supercontinuum generation[18] by a fs Yb:fiber laser free of the pulse picking and a bare fiber several cm in length[19-21] (Fig. 1a, Approach 3). Despite these progresses, the corresponding nonlinear fibers have a relatively small core (<12 μm) and do not support high-peak power coherent fiber supercontinuum generation by the pp-FCPA lasers (Table S1). In this context, our recent attempt using a pp-FCPA laser (Satsuma, Amplitude) and a large-core (15 μm) photonic crystal fiber (LMA-PM-15, NKT Photonics)[22] put us in a position to develop the alternative tuning accessory (Table S2).

Unfortunately, we found that the corresponding supercontinuum generating fiber inevitably suffered irreversible photodamage after ~100-hr of accumulative operation. This disruption prohibits the operation of the corresponding supercontinuum laser by a biologist (without extensive laser training). It is thus important to identify the nature of this long-term photodamage and then avoid it. We aimed to answer whether this photodamage was caused by airborne contaminant in a non-clean-room environment and/or high peak-intensity free-space coupling at two fiber end facets, which could be avoided by commercial photonic crystal fiber end-capping/termination with specific hole collapsing and beam expansion[23] (Fig. 1a, Approach 2), or other more complicated mechanisms.

*Experiment on two schemes of fiber supercontinuum*

Our custom-built coherent fiber supercontinuum source (Fig. 1a, Approach 3; Table S3, Scheme 1) enabled slide-free histochemistry[22], nonlinear optogenetics[24], and label-free imaging of extracellular vesicles[25]. The supercontinuum output along one principal axis of polarization-maintaining (PM) LMA-PM-15 fiber with a high polarization extinction ratio (PER) reproducibly exhibited the same spectrum (Fig. 1b, Scheme 1) for different cleaved 25-cm fiber pieces (Table S3), as asserted by a deterministic model[26] taking account of polarization effect[27]. However, each piece encountered a long-term photodamage after accumulative (not continuous) operation of 100±40 hr (20 pieces in total), resulting in gradually reduced (up to 10%) coupling efficiency not compensable by optical realignment along with narrowed spectral broadening and often degraded output beam quality. We observed this fiber photodamage in another polarized supercontinuum source[28], except for the use of a non-FCPA operated at 40 MHz as the master laser (Table S3, Scheme 2). Although Scheme 2 matched Scheme 1 in input peak intensity (Table S3), the resulting supercontinuum produced a broader spectrum due to the lower dispersion of the fiber (Fig. 1b, bottom). However, photodamage with gradually reduced coupling efficiency was found to occur in a shorter timeframe of 10±2 hr (Table S3), so that the stain-free histopathology had to replace the fiber daily to obtain reproduceable results[29]. This photodamage required replacement of the fiber with tedious optical realignments and thus limited the femtosecond biophotonics application of both schemes of supercontinuum source.

We identified one key difference between the two schemes. The photodamage in Scheme 2 was localized within 1-cm beyond the entrance end of the fiber, as re-cleaving of this length for a damaged 25-cm fiber piece would recover the fiber coupling efficiency and supercontinuum bandwidth. In contrast, the photodamage in Scheme 1 was relatively delocalized, as re-cleaving up to 10-cm length beyond the entrance end of a damaged 25-cm fiber piece was needed to recover the fiber coupling efficiency. The observed localization of fiber photodamage and reduced coupling efficiency over time are inconsistent with airborne contamination in a non-clean-room environment and/or high peak-intensity free-space coupling. The former would lead to rather sudden or random reduction of the coupling efficiency while the latter spatiotemporally similar photodamage for the two schemes. Thus, it is unlikely to mitigate the photodamage by specific fiber end-capping with mode expansion (Fig. 1a, Approach 2).

*Test on a third scheme of fiber supercontinuum*

The observed photodamage in Scheme 2 supports the interpretation based on the emergence of a photoscattering waveguide at the fiber entrance end[30,31] in the form of long-period fiber grating (LPFG)[32]. In this interpretation, input pulse propagating in the core mode beats with the copropagating pulse in a cladding mode after free-space-to-fiber coupling to produce the standing wave that writes and progressively strengthens a long-period fiber grating (LPFG). The period ($\Lambda$) of this LPFG is determined by the phase matching of $\Lambda = \lambda/[n_{co}(\lambda)-n_{cl}(\lambda)]$, where $\lambda$ is the central wavelength of the pulses while $n_{co}(\lambda)$ and $n_{cl}(\lambda)$ are the corresponding effective refractive index of the core mode and cladding mode, respectively. The pulses have broad bandwidths (~10 nm for 280 fs input and larger along the fiber for the core mode due to supercontinuum generation) from which the blue and red edges write slightly different grating periods and lead to the localized LPFG formation at the entrance end (because the superposition of the gratings from different wavelengths can be in phase for a limited length). The temporal walk-off between two pulses may also contribute to this localized LPFG formation.

For a given $\lambda$, the period $\Lambda$ of a circular fiber can be calculated from the dielectric structure of fiber cross section[32]. Similarly, $\Lambda$ of a photonic crystal fiber can be calculated from the pitch and hole sizes of fiber cross section (Fig. 1b, Inset) for the two schemes (Table S3), if $n_{cl}(\lambda)$ approximates the effective refractive index of the fundamental space filling mode[33]. The much larger $\Lambda$ in Scheme 1 as opposed to Scheme 2 is thus responsible for more delocalized fiber photodamage to approach similar LPFG strength (with dozens of periods) or loss of fiber coupling efficiency (10%), and slower LPFG formation via increased spectral broadening (supercontinuum generation) and/or pulse walk-off at longer fiber lengths. As a nontrivial prediction from this interpretation, the LPFG-based photodamage would disappear if the calculated $\Lambda$ approaches the total length of the supercontinuum-generating fiber (because the LPFG would function poorly with only one period).

To test this prediction, we developed a third scheme of supercontinuum generation using an ultra-large core silica photonic crystal fiber (LMA-PM-40-FUD, NKT Photonics) that approximates the doped DC-200/40-PZ-Yb fiber in a PP-FCPA laser[12], with a cross section of large-pitch small-hole lattice (Table S3, Scheme 3). The selection of a short fiber length (9.0 cm) not only avoided undesirable bending effect[34] or depolarization effect[27], but also approached the calculated $\Lambda$ from this fiber (Table S3). Without the fiber end-capping (Fig. 1a, Approach 2), the resulting supercontinuum source (Fig. 1b, Scheme 3) remained stable after >2000 hours of accumulative operation within 2 years in a regular (non-clean-room) optical laboratory. This test validates our LPFG-based interpretation of fiber photodamage. Besides the suppression of the LPFG photodamage, the large core size (40 μm) also scales up the peak power for tunable pulse generation (see below), just like that for non-dissipative[34] and dissipative soliton pulses[35].

*Fiber-optic nonlinear wavelength converter*

We next examined the dependence of supercontinuum spectrum (Table S3, Scheme 3) on $f$ of the master PP-FCPA laser at the same $E$ (i.e., $P/f$). The laser/input spectrum and $\tau$ was rather independent on $f$, so that the supercontinuum output retained similar spectrum across wide $f$ range of 2-10 MHz (Fig. 1c, top). This deterministic generation of coherent fiber supercontinuum is not surprising because the spectrum can be theoretically predicted if the spatiotemporal property of input laser pulse is known[26]. Similar $f$-independent spectra were obtained at lower $E$, so that the leftmost and rightmost spectral lobes may be filtered to generate compressed pulses[21] across 950-1110 nm, wherein they converge at 1030 nm (Fig. 1c, bottom). The observed $f$-independent supercontinuum generation resembles that of soliton generation[36].

Experimentally, collimated fiber supercontinuum output was aligned along the horizontal polarization by an achromatic half-wave plate to enter a pulse dispersion compensation unit (Fig. 2a) in the form of programmable pulse shaper (FemtoJock, Biophotonic Solutions), which was empowered by multiphoton intrapulse interference phase scan (MIIPS)[37] through a 128-pixel spatial light modulator (SLM)[38]. The pulse shaper spectrally selected a fixed-bandwidth window (~60 nm) inside the supercontinuum spectrum with a tunable central wavelength across 950-1110 nm after motorized rotation of the reflective grating of the pulse shaper to project this spectral window on the SLM[39]. For a pulse centered at $\lambda$ = 1030 nm without the spectral lobe filtering or at a detuned $\lambda$ (e.g., 1110 nm) with this filtering[21], the pulse shaper allowed compressing this pulse close to its transform limit $\tau$ (~60-fs FWHM or ~40-fs sech$^2$-shape)[37,39] and chirping/tuning the pulse to ~400 fs (Fig. 2b).

Optionally, the free-space output from the pulse shaper was recoupled into a 1-m low-dispersion Kagome hollow-core fiber patch cable (PMC-C-Yb-7C, GLOphotonics) by an achromatic lens of 75-mm focal length, with slightly $\lambda$-dependent efficiency of 76±3%. The weak birefringence intrinsic to the hollow-core fiber[40] allowed rotating the input polarization by a half-wave plate to maximize the PER of fiber-delivered output to 10-20, depending on the bending state of the fiber. The spectrum and spatial beam profile of fiber-delivered output after the collimation by an achromatic lens (75-mm focal length) approximated those of the free-space input before the fiber, while the small pulse duration of free-space input was largely retained (Fig. 2b). At the cost of 24% lower $P$ or $E$ and slightly degraded PER, the fiber pulse delivery gains several advantages over free-space pulse delivery: i) simple fiber telecommunication connection and disconnection allows easy switching among different optical fiber-coupled application modules, i.e. sharing the fiber delivered output among these modules (Fig. 2a); ii) fiber delivery of energetic pulses is safer than free-space delivery for operators without extensive laser training; iii) the optimal fiber recoupling condition of endless single-mode fiber supercontinuum[41,42] is independent on $\lambda$ (i.e. rotation of the grating in the pulse shaper), which can be useful to monitor and correct the misalignment of the pulse shaper itself[38] in portable application of this tunable source (beyond an environmentally controlled laboratory).

The SLM-based pulse shaper is not necessary for the tunable fiber supercontinuum source with or without hollow-core fiber delivery when only tunable-$\tau$ pulse generation (rather than arbitrary pulse shaping[39]) is needed. We tested a more cost-effective alternate of single-prism pulse compressor (BOA-1050, Swamp Optics) and generated similar tunable-$\lambda$ ~40-fs (sech$^2$) pulse by motorized rotation of the prism and the linear motion of a back retroreflector that varies group delay dispersion (GDD) (Table S4), indicating that the chirp of this fixed-bandwidth pulse is largely linear. Due to the fiber input (supercontinuum generation) and optional fiber output (dispersion-free pulse transmission through the hollow-core fiber) of the dispersion compensation unit, we term the whole device a fiber-optic nonlinear wavelength converter (FNWC) (Fig. 2a, Table S4). Our FNWC may be generalized to other pp-FCPA lasers (Table S1) with $f$-independent emission spectrum (Fig. 1c, top). In contrast to commercial alternatives such as an OPA or OPO, FNWC can independently and widely tune $\lambda$, $f$, and $\tau$ (Tables S4, S5).

There is room to further improve the existing FNWC technology. Broader bandwidths of fiber supercontinuum generation at high input powers may be possible if the multimodal behavior at longer wavelengths (>1120 nm)[12] and the bleed-through of long-wavelength tail of supercontinuum into fiber anomalous dispersion regime (>1250 nm, Table S3) would not degrade single-mode coherent supercontinuum generation. Also, silica photonic crystal fibers with even larger core, e.g., 100 μm (SC-1500/100-Si-ROD, NKT Photonics)[43], may further increase the peak power for supercontinuum generation and the resulting FNWC output while restrict the LPFG-based photodamage. Finally, the improvement of hollow-core delivery fibers on single-mode low-loss transmission[44], bending tolerance, and polarization maintaining may continue.

The unique fiber delivery of spectrally filtered fiber supercontinuum pulses excels at user-friendly and cost-effective operation. First, for pulse parameters ($\lambda$, $f$, and $\tau$) of choice, spectrally monitoring the corresponding deterministically generated fiber supercontinuum ensures day-to-day reproducible optical alignment before the FNWC (Fig. 1a). Second, for a pre-selected spectrum of fiber supercontinuum, monitoring the corresponding fiber delivery output spectrum, power (and plausibly PER), and modal content[45] ensures day-to-day reproducible optical alignment before the application modules (Fig. 2a). Third, the fiber-optic telecommunication-based connection and disconnection of the delivery fiber not only ensures the benefit of laser-microscope alignment decoupling[29], but also enables simple switching or sharing of an integrated pp-FCPA-FNWC laser among different microscopes or applications (Fig. 2a).

*Utility in high-performance label-free multiphoton microscopy*

To test the FNWC in laser-scanning multiphoton microscopy[46], which is known for overall good performance in 3D sectioning ability, molecular sensitivity/specificity (via fluorescence), and image content (e.g., field of view, spatial resolution, and depth). We replaced the supercontinuum source (Table S3, Scheme 1) of our simultaneous label-free autofluorescence multi-harmonic (SLAM) microscope[47] with the FNWC (Table S3, Scheme 3), and collimated the fiber-delivered pulses by an achromatic lens as a free-space beam input to the microscope (Fig. 2a). By operating the FNWC at the optimized illumination of SLAM imaging that integrates 4 modalities of two- and three-photon excited auto-fluorescence and harmonics (2PAF, SHG, 3PAF, and THG; see Table 1), we reliably visualize live samples such as ex vivo rodent tissue (Fig. S1). The plausible high-order mode coupling and related side-pulse generation (Fig. 2b) in the delivery fiber did not degrade the imaging performance, as also demonstrated in multiphoton microscopy with Kagome hollow-core fiber delivery of Ti:sapphire laser pulses[48]. However, the FNWC retains stable output after >2000 hr (and counting) of cumulative operation without replacing the supercontinuum generating photonic crystal fiber, overcoming a critical limitation of the original SLAM microscope. Beyond regular SLAM imaging with long-term stability, FNWC can adapt to evolving research needs such as high temporal-resolution intravital imaging and high-throughput fluorescence lifetime imaging microscopy (FLIM), by tuning the illumination (i.e., repetition rate from 10 MHz to 5 MHz) and building an extended SLAM microscope (eSLAM) with a fiber-coupled input (Fig. 2a). We replaced the slow flexible optical scanner and 4 photon-counting photomultipliers in the SLAM microscope with a fast inflexible scanner and 4 analog-detection photomultipliers that enabled FLIM via single-photon peak event detection[49] (Table 1). The mode-locking electronic signal was used as the master clock to synchronize optical scanning and subsequent signal acquisition.

The increased speed of eSLAM over regular SLAM imaging lowered the excitation cycle to the minimum of 1 pulse per pixel per frame (Table 1), and thus limited the signal-to-noise ratio (SNR) in the intravital imaging of a mouse skin flap (Video S1) with THG-visible flowing blood cells (Fig. 3a, arrows) along with SHG-visible collagen fibers and periodic sarcomeres along muscle myofibrils[50] (Fig. 3a, arrowhead). This low SNR is often encountered in real-time nonlinear optical imaging free of labeling and phototoxicity. With the advent of modern machine learning models such as DeepCAD-RT[51], the SNR can be improved considerably (Video S2), but at the cost of the ability to track individual blood cells (Fig. 3b, arrows) and resolve the periodic sarcomeres (Fig. 3b, arrowhead). In sharp contrast, another machine learning model known as UDVD[52] not only improves the SNR considerably but also recovers this ability (Fig. 3c, arrows and arrowheads; Video S3), indicating its better suitability for our intravital imaging of dynamically moving samples. For weaker 2PAF/3PAF signals collected simultaneously, UDVD reveals 2PAF-visible stromal cells (Fig. 3d, arrows) and 3PAF-visible lipids (Fig. 3d, stars) barely discernable in the raw data (Fig. 3e), exhibiting a larger SNR improvement in comparison to the SHG/THG signals (Fig. S2). Also, at two different instances of imaging, UDVD unambiguously reveals the presence of intracellular 2PAF, 3PAF, and THG signals in different parts of single biconcave disk-shaped blood cells (Figs. 3f, 3g, arrows; Fig. S3 vs. Video S1), which can be confirmed by imaging fresh blood smear sample (Fig. S4). In fact, UDVD benefits all instances of time-lapse eSLAM imaging despite the inevitable sample movement during the imaging (Video S3 vs. Video S1).

To demonstrate the high throughput of eSLAM imaging and its built-in FLIM ability in live-tissue pathology, we collected THG (Fig. S5a), SHG (Fig. S5b), 3PAF intensity (Fig. S5c), 3PAF lifetime (Fig. S5d), 2PAF intensity (Fig. S5e), and 2PAF lifetime (Fig. S5f) images from a large (1-mm$^2$) area of *ex vivo* mouse kidney tissue in 30 min. All images involved a 5×5 mosaic of field-of-views with an overlapping factor of 20%, which was enabled by an automatic mechanical stage. Among these images, only 2PAF lifetime unambiguously reveals the large-scale vasculature expected from vital kidney tissue that has been visualized at a typical field-of-view of 250×250 μm$^2$ [53] (Fig. 3h, Fig. S5f). It should be noted that some elongated patterns of punctuated points in the 2PAF intensity image (Fig. S5e, arrows) do not co-register with red-colored vasculature in the 2PAF lifetime image (Fig. S5f). Similar large-scale data in a 3D volume were obtained in 12 min to reveal the depth-resolved vasculature (Video S4). It is conceivable that this high content imaging by FLIM-included eSLAM may help pathologists to diagnose diseases from fresh core biopsies or surgical specimens (optical biopsy).

**Discussion**

Our FNWC is not limited to label-free imaging. Due to the relatively high peak-power afforded by this device, its deficiency in two-photon excitation of common fluorophores below 950 nm may be compensated by three-photon excitation across 950-1110 nm. More importantly, the tunable aspect of FNWC will enable fast prototyping or optimization of imaging condition not available from alternative lasers[54]. For multiphoton microscopy with photon order $n$ (>1 integer), the signal generation rate scales with $P^n/(f\tau)^{n-1}$. Thus, a combined low-$f$ and short-$\tau$ excitation condition, i.e., a high duty-cycle inverse $(f\tau)^{-1}$, would enhance the signal at a given $P$, which is limited by laser safety of American National Standards Institute (ANSI). However, one well-known photodamage mechanism also scales with $P^r/(f\tau)^{r-1}$, in which the nonlinear order $r$ lies between 2 and 3[55]. Given a two-photon signal of interest ($n = 2$), the mitigation of this highly nonlinear ($2 < r < 3$) photodamage demands a low duty-cycle inverse $(f\tau)^{-1}$ because $n < r$. On the other hand, there exists another popular photodamage mechanism that includes two-photon absorption-induced photochemical damage ($r = 2$)[56] and one-photon absorption-induced photothermal damage ($r = 1$)[57]. Because $n \geq r$, the mitigation of this low-$r$ photodamage demands a high duty-cycle inverse. Thus, the flexibility in $f$ and $\tau$ is needed to optimize the signal-to-photodamage ratio for two-photon microscopy, depending on specific biological samples and photodamage mechanisms. With the advent of modern machine learning models such as UDVD and DeepCAD-RT, this optimization should emphasize more on a low photodamage than a high SNR and can be performed user-friendlily and cost-effectively by our tunable FNWC with unique fiber delivery of spectrally filtered fiber supercontinuum pulses. For free-space output, the stable supercontinuum generation by FNWC allows programmable label-free contrast generation for gentle multiphoton microscopy[29].

The key enabling feature of FNWC is the surprising suppression of the long-term fiber photodamage in coherent supercontinuum generation using a photonic crystal fiber with large-pitch small-hole lattice. With this innovation, one laser source can serve both the regular SLAM and eSLAM microscope, which complement each other in different imaging applications (Table 1). Once the corresponding optics is pre-aligned, the switch between the two types of imaging can be done by simple connection and disconnection of optical telecommunication within seconds/minutes (Fig. 2a, lower right), without any effort of optical realignment. Meanwhile, reproducible laser operation can be ensured by inline monitoring spectrometers and power meter (Fig. 1a, Fig. 2a). This unique adaptation to evolving research needs may be extended to non-imaging applications of femtosecond biophotonics, e.g., precision surgery, laser tweezer, and optogenetics[39], to address the unmet needs of: i) widely and independently tuned in $\lambda$, $f$, and $\tau$ with sufficient P or E; ii) extensive use of optical fibers robust against environmental perturbations that permits portable access to tunable ultrafast laser technology outside an environmentally controlled laboratory; and iii) optical fiber-delivered output that allows safe access to high-irradiance laser pulses by diverse users working in real-world situations (but without extensive laser training).

Finally, we note that various ultrashort pulses have been similarly generated by nonlinear compression based on self-phase modulation in the forms of gas cells[58], gas-filled hollow-core fibers[59], and multi-plates[60]. Also, optical pulses like those present in this study have been produced by gain-managed nonlinear amplification[61], which has been employed in multiphoton microscopy[62]. It is our hope that the reported FNWC, with its rather unique tunability in pulse repetition rate, will further expands femtosecond biophotonics to benefit field biologists, neuroscientists, veterinarians, surgeons, and pathologists.

## Methods

### Animal tissue

All animal procedures were conducted in accordance with protocols approved by the Illinois Institutional Animal Care and Use Committee at the University of Illinois at Urbana-Champaign. Kidney was excised from a rabbit (*Oryctolagus cuniculus*, Charles River Laboratories, Wilmington, MA), submerged in sterile $Ca^{2+}/Mg^{2+}$-free 0.1 μm filter-sterilized PBS (pH 7.0 – 7.2), and washed from blood by changing the PBS solution. The kidney was manually sliced in sterile tissue culture dish kept on ice. Individual tissue slices were then placed onto uncoated 35 mm imaging dishes with No. 0 coverslip and 20 mm glass diameter (MatTek, #P35G-0-20-C). The slices were incubated in 500 μL FluoroBright™ DMEM (ThermoFisher Scientific, #A1896701) supplemented with 10% FBS, 1% PSA, and 4 mM L-Glutamine solutions. Mice (C57BL/6J, Jackson Laboratory) were used to obtain *ex vivo* kidney samples and blood smear samples, which were imaged directly behind a coverslip in inverted SLAM/eSLAM microscope without solution-based preparation. Incision to expose mouse mammary tissue and externalization of the skin flap was performed under isoflurane anesthesia. The skin flap was placed on a large coverslip and imaged by inverted eSLAM microscope, while the mouse was anesthetized with 1% isoflurane mixed with $O_2$ at a flow rate of 1 L/min. Throughout the imaging, a heating blanket was used to maintain the physiological temperature. Imaging was limited to 3 h duration, after which the mouse was euthanized.

### Self-supervised denoising models

Two self-supervised denoising models, i.e., DeepCAD-RT and UDVD, were used for eSLAM video denoising. Individual channels (SHG, THG, 2PAF, and 3PAF) of the low-SNR videos were used to train UDVD or DeepCAD-RT separately for 100 epochs. To quantify the noise level for the whole video without noise-free ground truth, we defined signal-to-noise ratio (SNR) as μ/σ, where μ is the mean pixel value and σ is the corresponding standard deviation. SNR was measured across all frames in the original and denoised videos (Fig. S2). The machine learning was conducted on a workstation computer equipped with a central processing unit (CPU) (Xeon W-2195, Intel), four graphics processing units (GPUs) (RTX 8000, Nvidia), and 256 gigabytes of memory. The workstation operates on the Ubuntu system (version 18.04). DL-based video denoising and data analysis were conducted using Python (version 3.9). PyTorch (version 1.11.0) was used during the implementation of the denoising models. Scikit-learn (version 0.23.2) was used for the calculation of evaluation metrics. Plots were generated using Matplotlib (version 3.2.2) and Seaborn (version 0.11.0). Other Python libraries including Numpy (version 1.19.1), Pandas (version 1.1.2), and SciPy (version 1.5.2) were used to assist data analysis.

## Author contributions

G.W. and H.T. conceived the idea. G.W. and H.T. conducted related experiments. G.W., J.S., J.E.S., and R.R.I. performed data analysis. G.W., J.S., and H.T. drafted the manuscript. H.T. reviewed and edited the manuscript with inputs from all authors.

## Competing interests

Haohua Tu is in discussion with the Office of Technology Management at the University of Illinois at Urbana-Champaign on commercial potential of the developed technology. Other authors declare no competing interests.

## Acknowledgments

The authors thank Stephen A. Boppart for sharing his laboratory, providing laboratory equipment resources, and for mentoring Jindou Shi, Rishyashring R. Iyer, and Janet E. Sorrells, and thank Edita Aksamitiene and Eric J. Chaney for preparing biological samples. H.T. acknowledges the financial support from the National Institutes of Health, U.S. Department of Health and Human Services (R01 CA241618). J.E.S. and R.R.I were supported by NIBIB/NIH under Award Number T32EB019944.

## Data availability

Data underlying the results presented in this paper are within the manuscript and Supplementary files.

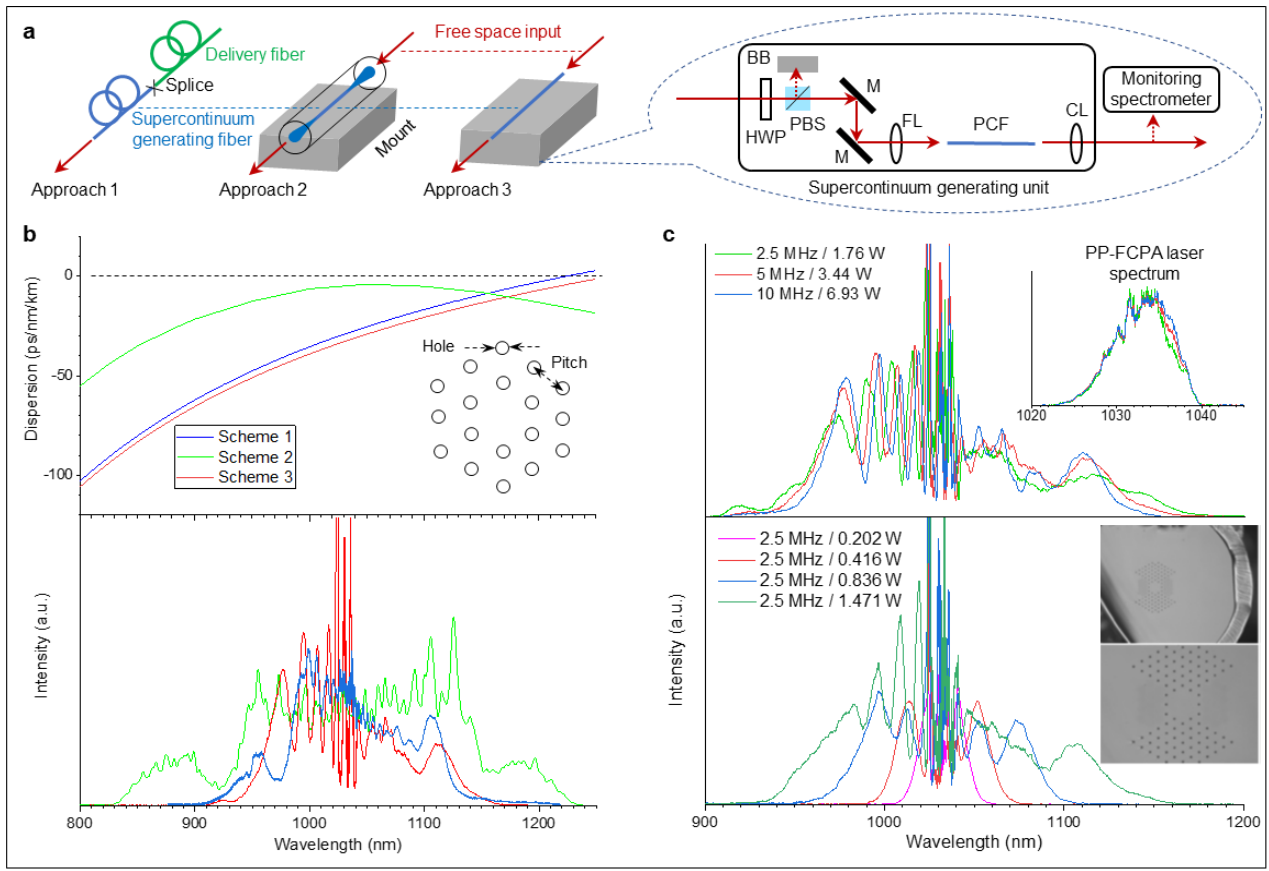

**Fig. 1**. **a** Three general approaches for fiber supercontinuum generation: all-fiber splice often used in commercial supercontinuum lasers (Approach 1), commercial enclosed device with fiber end capping and mode expansion as an add-on nonlinear wavelength converter for a Ti:sapphire oscillator (Approach 2), and mounted bare (polarization-maintaining) fiber for coherent fiber supercontinuum generation by a fs Yb:fiber laser (Approach 3). PP-FCPA – pulse-picked fiber chirped pulse amplifier, BB – beam blocker, HWP – halfwave plate, PBS – polarizing beam splitter, M – mirror, FL – focusing lens, PCF – photonic crystal fiber, CL – collimating lens; **b** three schemes of polarized coherent fiber supercontinuum generation under study with wavelength-dependent dispersion of photonic crystal fibers indicative of the restriction of supercontinuum generation to fiber normal dispersion regimes (top) with cross-sectional image of photonic crystal fibers indicative of pitch and hole sizes (inset), and corresponding spectra of supercontinuum outputs (bottom); **c** Output spectra at different *f* but the same *E* for Scheme 3 (top) in comparison with input spectra of source laser (inset), and output spectra at different *E* but the same *f* for Scheme 3 (bottom) with cross sectional images of the supercontinuum generating fiber (inset).

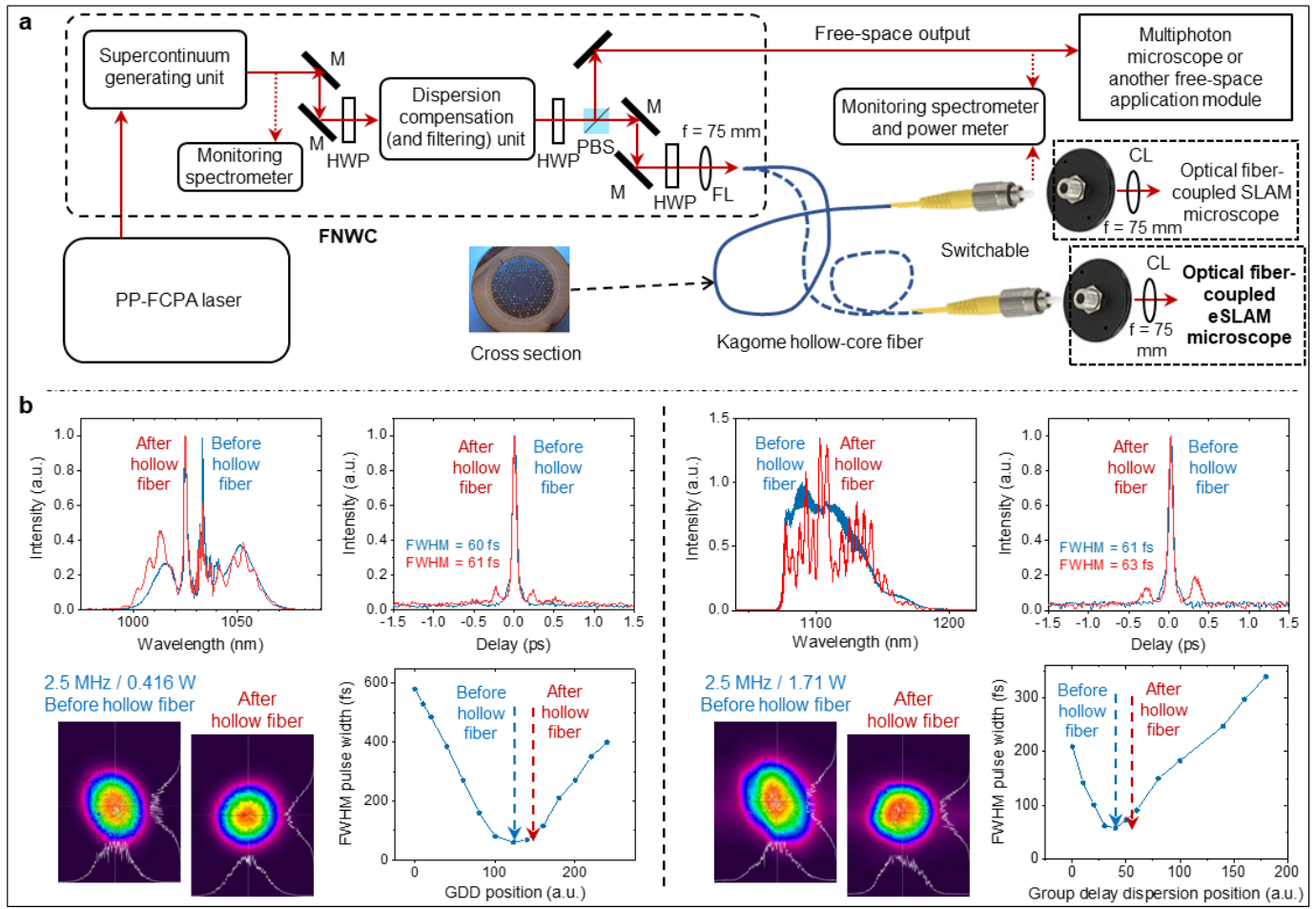

**Fig. 2**. **a** Schematics of fiber-optic nonlinear wavelength converter (FNWC) and related optical components for femtosecond biophotonics switchable between different microscopes (or applications) by fiber-optic telecommunication connection and disconnection. PP-FCPA – pulse-picked fiber chirped pulse amplifier, BB – beam blocker, M – mirror, HWP – halfwave plate, PBS – polarizing beam splitter, FL – focusing lens, CL – collimating lens; **b** FNWC output spectrum (1030-nm central wavelength without filtering the supercontinuum), pulse width, spatial mode/profile, and full width at half maximum (FWHM) pulse width versus group delay dispersion (GDD) position before and after 1-m Kagome hollow-core fiber (left), in compassion to FNWC output spectrum (1110-nm central wavelength from filtered supercontinuum), pulse width, spatial mode/profile, and FWHM pulse width versus GDD position before and after 1-m Kagome hollow-core fiber (right).

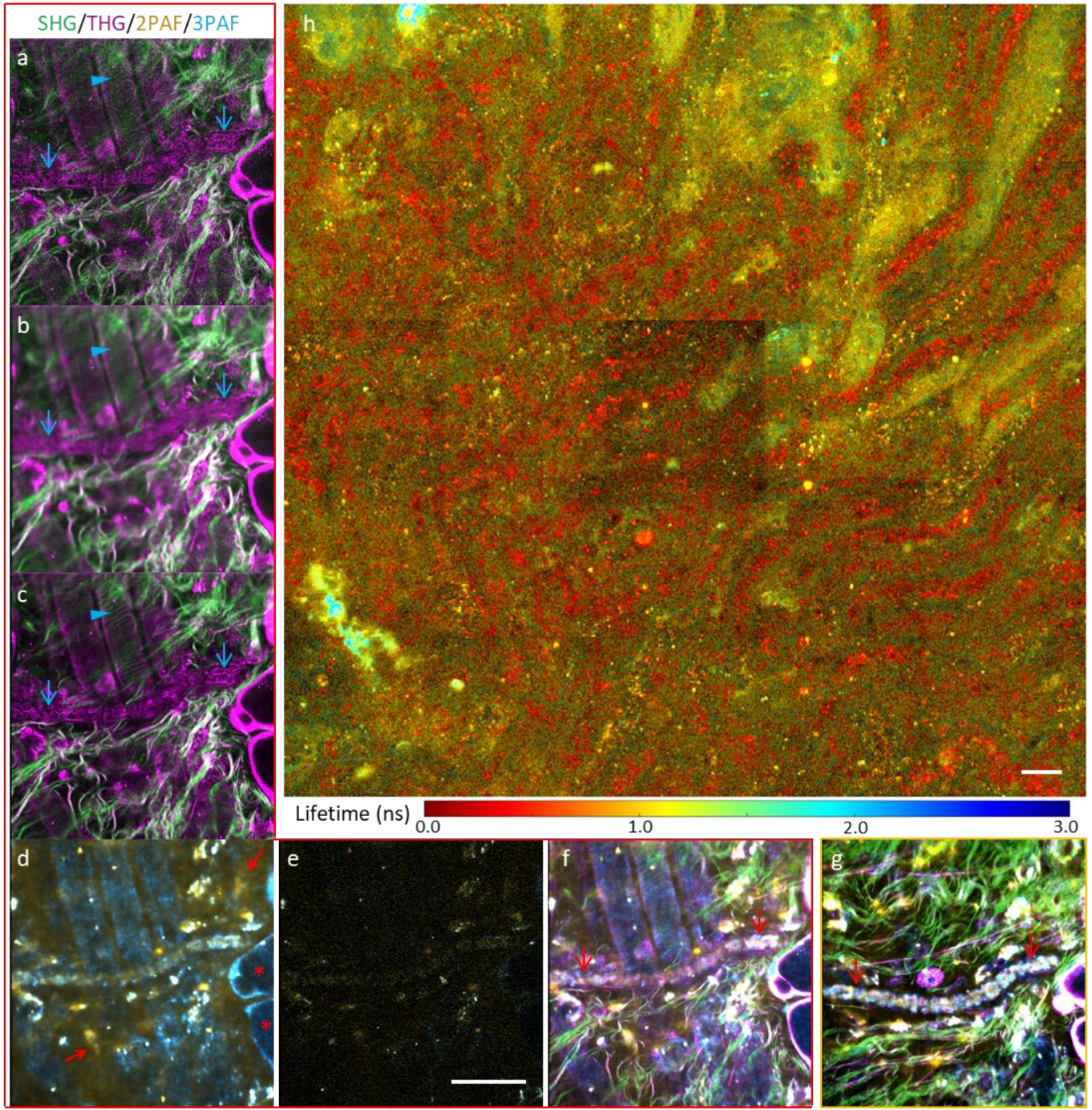

**Fig. 3. FLIM-empowered eSLAM imaging of unlabeled live specimens by FNWC.** Scale bar: 50 μm. **a-g** time-lapse intravital imaging of a surgically opened mouse skin flap at one instance, showing flowing blood cells in a blood vessel (cyan arrows) and periodic sarcomeres along muscle myofibrils (arrowhead) in raw SHG/THG data **a** which are blurred in DeepCAD-RT-denoised data with a better overall SNR **b** but recovered in UDVD-denoised data with a better overall SNR **c**; the UDVD-denoised data **d** also reveal stromal cells (red arrows) and lipids (stars) barely visible in raw 2PAF/3PAF data **e**, resulting in a composite 4-color image with discernible blood cells at one instance **f** that can be compared with the similar image at a different instance **g** (see Video S3). **h** 2PAF lifetime (FLIM) image of *ex vivo* mouse kidney tissue over a 5×5 mosaic of field-of-views (1 mm² total area) that shows red-colored large-scale vasculature with a fluorescence lifetime of <0.6 ns.

**Table 1. Complementary features of regular SLAM imaging and eSLAM imaging that share one FNWC.**

|  | Regular SLAM | eSLAM |
| --- | --- | --- |
| Pulse repetition rate (average power) | 10 MHz (≤17 mW*) on sample | 5 MHz (≤17 mW*) on sample |
| Optical scanner; fast-axis line rate | Galvo-Galvo (6215 H, Cambridge Technology); up to 350 Hz | Resonant (SC30, Electro-Optical Products) and Galvo (GVS011, Thorlabs); 1592 Hz |
| Photodetection mode | Photon counting | Analog sampling (2 GHz for 2PF/3PF; 125 MHz for SHG/THG) |
| PMT – 2PAF, quantum efficiency | H7421-40 (Hamamatsu), 31.6% | H7422A-40 (Hamamatsu), 41.4% |
| PMT – 3PAF, quantum efficiency | H7421-40 (Hamamatsu), 31.8% | H7422A-40 (Hamamatsu), 42.1% |
| PMT – SHG, quantum efficiency | H7421-40 (Hamamatsu), 33.4% | H10721-20 (Hamamatsu), 16.8% |
| PMT – THG, quantum efficiency | H7421-40 (Hamamatsu), 20.4% | H10721-210 (Hamamatsu), 42.4% |
| Frame size (field-of-view) | 700×700 pixels (≤300 × 300 µm$^2$) | 1024×1024 pixels (250 × 250 µm$^2$) |
| Pixel dwell time (pulses/pixel/frame) | 2-10 µs (20-100) | 0.2 µs (1) |
| Frame illumination/acquisition time | 1-5 s | 0.33/1.37 s |
| Raw data acquisition for real-time display and storage | Enabled by a regular CPU | Enabled by a GPU (GeForce RTX 2080, NVIDIA) |
| Strengths | Low detection noise and flexible optical scanning | High temporal resolution with FLIM capability |
| Weaknesses | Low temporal resolution that would be worsened by FLIM | Large detection noise and inflexible optical scanning |
| Suitable applications | Quantitative live-cell imaging for drug discovery, label-free imaging with weak signals, small-scale optical biopsy, etc. | Imaging dynamically moving live samples, label-free imaging with moderate signals, labeled imaging, large-scale optical biopsy, etc. |

* - limited by phototoxicity. Common features: illumination band - 1110±30 nm; pulse width on sample - 60 fs (FWHM); microscope objective - UAPON40XW340 (Olympus), NA 1.15 water immersion.

**Supplemental document**

**Video S1.** Time-lapse intravital eSLAM imaging of mouse skin flap without image denoising.

**Video S2.** Time-lapse intravital eSLAM imaging of the mouse skin flap denoised by DeepCAD-RT.

**Video S3.** Time-lapse intravital eSLAM imaging of the mouse skin flap denoised by UDVD.

**Video S4.** Depth-resolved eSLAM imaging of *ex vivo* mouse kidney with SHG/green against THG/magenta contrasts (upper left), 2PAF/yellow aganist 3PAF/cyan contrasts (lower left), 2PAF lifetime contrast (upper right), and 3PAF lifetime contrast (lower right).

**Table S1. Representative commercial Yb-based pp-FCPA lasers and optional OPA accessories**

| Model (vendor) | $\lambda$, $f$, and $\tau$ (low-bound) * | $P$ | $\lambda$, $f$, and $\tau$ from optional OPA |
|---|---|---|---|
| Satsuma (Amplitude) | 1030 nm, 0.5-40 MHz, 300 fs | 5/10/20 W | Niji: 257-4000 nm, 0.5-2 MHz, 50 fs |
| Y-Fi (Thorlabs) | 1035 nm, 1-10 MHz, 220 fs | 3-20 W | Yi-F OPA: 1275-1800 nm, 1-3 MHz, 100 fs |
| Spirit 1030 (Newport) | 1030 nm, 1-30 MHz, 400 fs | 70/100/140 W (water cooled) | NOPA-VISIR: 650-900 & 1200-2500 nm, 1-4.3 MHz, 70 fs |
| Monaco (Coherent) | 1035 nm, 1-50 MHz, 350 fs | 40/60 W (water cooled) | Opera-F: 650-900 & 1200-2500 nm, 1-4 MHz, 70 fs |
| FemtoFibe vario 1030 HP (Toptica) | 1030 nm, 1-10 MHz, 250 fs | 8 W | N.A. |
| Impulse (Clark-MXR) | 1030 nm, 2-25 MHz, 250 fs | 20 W | N.A. |
| BlueCut (Menlosystem) | 1030 nm, 1-10 MHz, 400 fs | 10 W | N.A. |
| FCPA-DE (IMRA) | 1045 nm, 1-5 MHz, 400 fs | 20 W | N.A. |

*Pulse width of sech2-profile that can be tuned (stretched) to several ps (up-bound) by a grating compressor.

**Table S2. Representative supercontinuum generation in bulk media and photonic crystal fibers**

| Report | Input $\lambda$, $f$, and $\tau$, respectively | Coupled $P$ (or $E$) | Interactive medium (diameter, length) | Feature/comment |
|---|---|---|---|---|
| 14 | 530 nm, Q-switch, 4-8 ps | (5mJ) | Bulk glass (1.2 mm, 2-1000 mm) | Discovery of supercontinuum generation under self-focusing |
| 15 | 800 nm, 0.25 MHz, 170 fs | 0.25 W (1 µJ) | Sapphire plate (~20 µm, <500 µm) | Bulk supercontinuum generation employed in commercial OPA |
| 16 | 770 nm, ~80 MHz, 100 fs | (0.2 nJ) | Photonic crystal fiber (1.7 µm, 10 cm) | Widely accessible single-mode fiber supercontinuum generation |
| 17 | 1070 nm, 40 MHz, 3.3 ps | 1.5 W | Photonic crystal fiber (4 µm, 5-65 m) | Commercial all-fiber supercontinuum generation for the broadest spectra |
| 22 | 1030 nm, 10 MHz, 300 fs | 1.8 W | Photonic crystal fiber (15 µm, 25 cm) | High peak-power coherent fiber supercontinuum generation |

**Table S3. Three schemes of polarized coherent fiber supercontinuum in this study**

| | Scheme 1 | Scheme 2 | Scheme 3 |
|---|---|---|---|
| Master laser (vendor) | Satsuma 10W (Amplitude) | Satsuma 10W (Amplitude) | |
| Input ($\lambda$, $f$, and $\tau$) * | 1030 nm, 10 MHz, 280 fs | 1031 nm, 40 MHz, 290 fs | 1031 nm, 5 MHz, 290 fs |
| Photonic crystal fiber (vendor) | LMA-PM-15 (NKT Photonics) | NL-1050-NEG-PM (custom, NKT Photonics) | LMA-PM-40-FUD (NKT Photonics) |
| Core diameter (mode field diameter) | 14.8 µm (12.6 µm @1064 nm) | 2.4 µm (2.2 µm @1064 nm) | 40 µm (32 µm @1064 nm) |
| Hole/pitch size; cladding diameter | 4.9/9.8 µm; 230 µm | 0.65/1.44 µm; 125 µm | 7/26 µm; 450 µm |
| Fiber zero dispersion wavelength | 1210 nm | Nonexistent | 1260 nm |
| Focusing length, lens (vendor) | 18.4 mm, C280TMD-B (Thorlabs) | 3.1 mm, C330TME-B (Thorlabs) | 50 mm, AC127-050-B (Thorlabs) |
| Coupled efficiency | 75% | 70% | 79% |
| Coupling (output) $P$ | 1.2 W | 0.22 W | 3.44 W |
| Input peak intensity** | 5 TW/cm$^2$ | 5 TW/cm$^2$ | 6 TW/cm$^2$ |
| Collimating parabolic mirror | focal length 25 mm | focal length 10 mm | focal length 50 mm |
| Output PER | >50 | >50 | >30 |
| Long-term fiber photodamage | present after 100±40 hrs accumulative operation | present after 10±2 hrs accumulative operation | absent after >2000 hrs accumulative operation |
| Number of fiber pieces | 18 | 7 | 2 |
| Fiber length | 25 cm | 25 cm | 9.0 cm |
| Localization of photodamage | <10 cm beyond fiber entrance end | <1 cm beyond fiber entrance end | Not observed |
| **Estimated $\Lambda$ of LPFG** | **1 mm** | **80 µm** | **~9 cm** |

*Pulse width of sech2-shape from two similar lasers (Satsuma 10W, Amplitude) with a beam diameter of ~1.8 mm. **Assuming no pulse broadening from laser exit to coupling photonic crystal fiber.

**Table S4. Comparison of FNWC with OPA as accessory for pp-FCPA lasers**

| Accessory | FNWC (SLM or prism compression) [*] | OPA |
|---|---|---|
| $\lambda$ - tunable range (nm) | 950-1110 | Wide |
| $f$ - variable range (MHz) | 1-10 | <5 (typically fixed) |
| $\tau$ - tunable range (fs) | 40-400 | ~70 (typically not tuned) |
| $P$ in mW ($E$ in nJ) | 20-200 (20) | up to 1000 (up to 250) |
| Independently tuned $\lambda$, $f$, and $\tau$ | demonstrated | not demonstrated |
| Fiber delivered output | demonstrated | not demonstrated |
| Portability/troubleshooting | possible/simple | limited/complex |

[*] Performance is from Satsuma 10W (Amplitude) and can be further improved for f and P using Satsuma 20W.

**Table S5. Comparison of integrated pp-FCPA-FNWC laser with tunable solid-state femtosecond lasers.**

| | pp-FCPA-FNWC laser | Tunable solid-state femtosecond lasers |
|---|---|---|
| Tunable $\lambda$ range (nm) | 950-1110 | 690-1020 (Ti:sapphire) or 690-1300 (OPO) |
| Other tunable pulse parameters | $\tau$ (down to ~40 fs) and $f$ (1-10 MHz in this study) | $\tau$ (typically, down to ~100 fs); $f$ of ~80 MHz typically not tunable |
| $E$ and $P$ | high and moderate, respectively | moderate and high, respectively |
| Dependence of $f$ on $\lambda$ | no | yes (even though small) |
| Beam pointing stability | ensured by endless single-mode supercontinuum (intrinsic) | ensured by feedback beam-pointing correction (extrinsic) |
| Cooling | air cooling sufficient | water cooling required |
| Maintenance cost | low | high |
| Comments on multiphoton excitation of common fluorophores | missing 2-photon excitation across 690-950 nm can be recovered by 3-photon excitation across 690-950 nm | often sufficient by 2-photon excitation across 690-1020 nm, except for some with red-shifted emission across 1020-1300 nm |

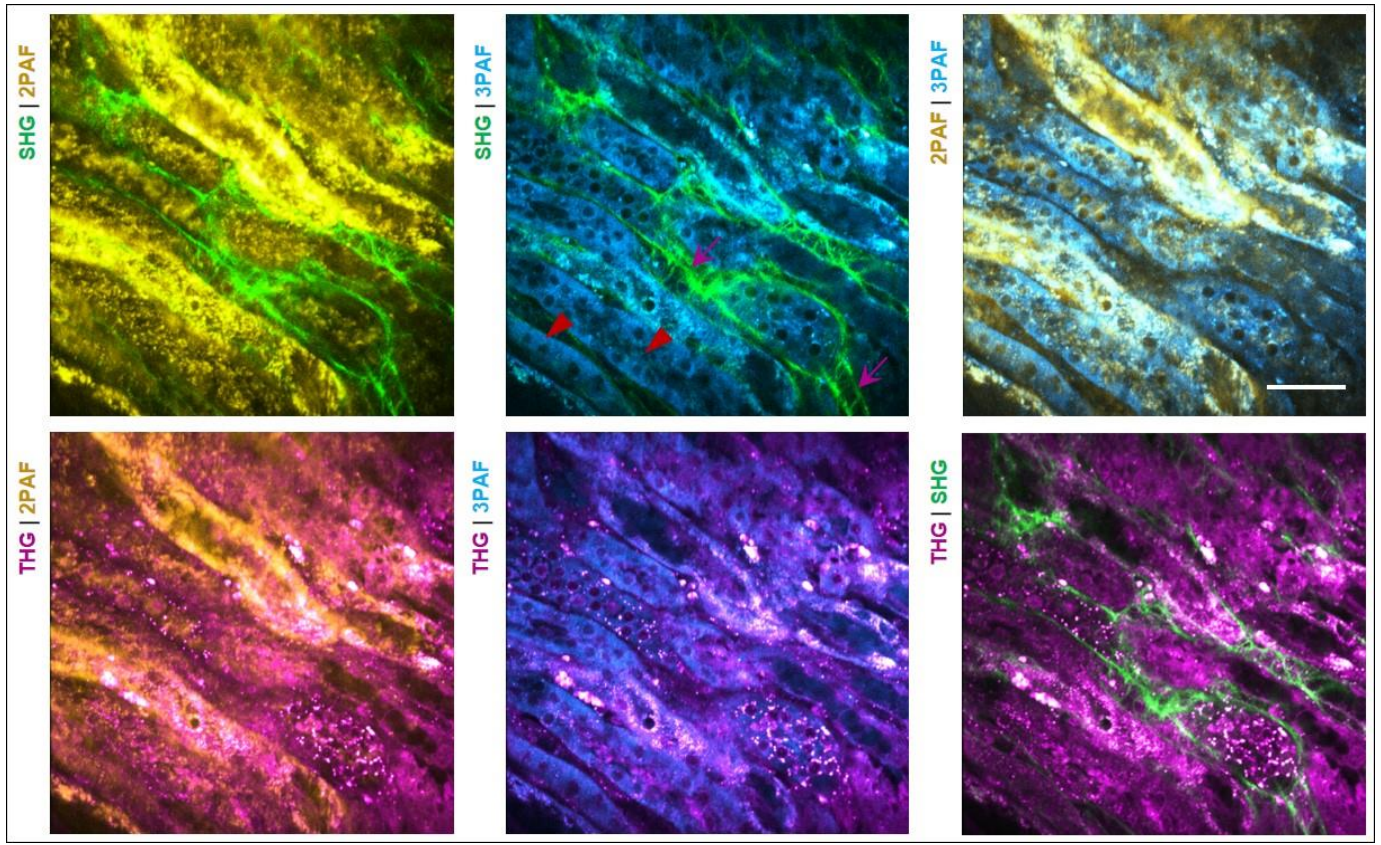

Fig. S1. Label-free regular SLAM images (17-s total acquisition time) of ex vivo rabbit kidney tissue excited by optical fiber-delivered FNWC output, with colored contrasts of second-harmonic generation (SHG), third-harmonic generation (THG), two-photon-excited auto-fluorescence (2PAF), and three-photon-excited auto-fluorescence (3PAF) showing live cells (arrowheads) and extracellular matrix (arrows). Scale bar: 50 µm.

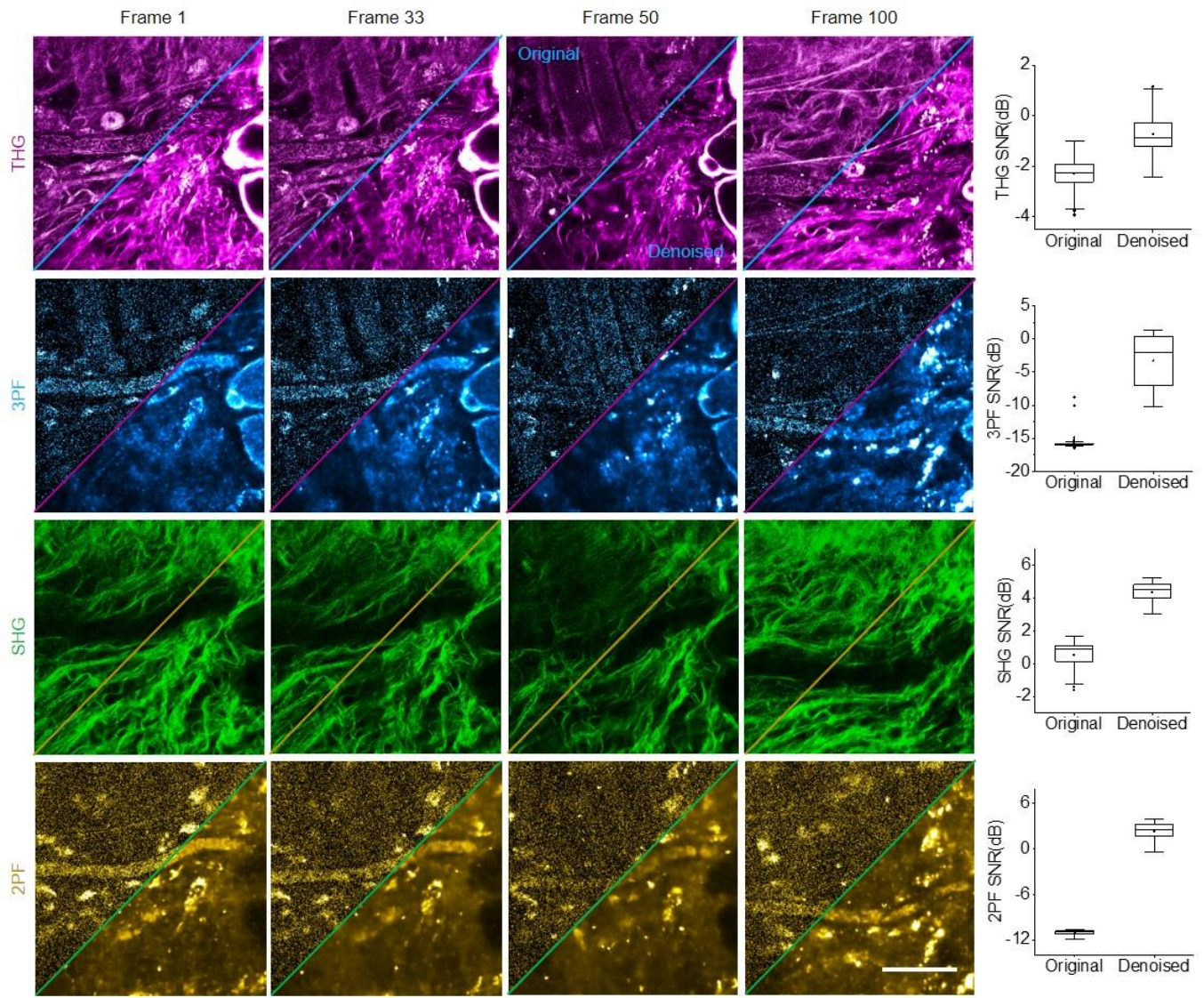

Fig. S2. Time-lapse intravital eSLAM imaging of mouse skin flap across the modalities of THG, 3PAF, SHG, and 2PAF without (upper left) and with UDVD denoising (lower right). The box and whisker plots show the corresponding SNR improvement (right panel). Scale bar: 50 μm.

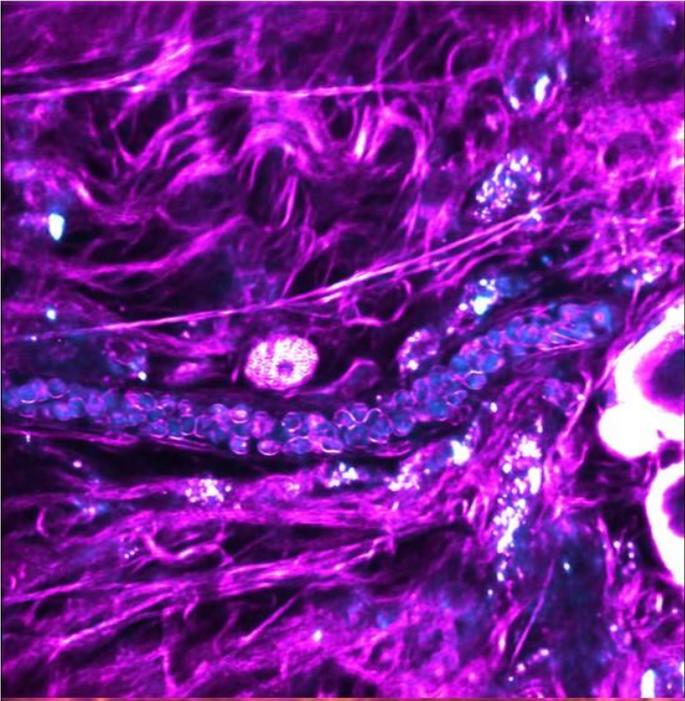 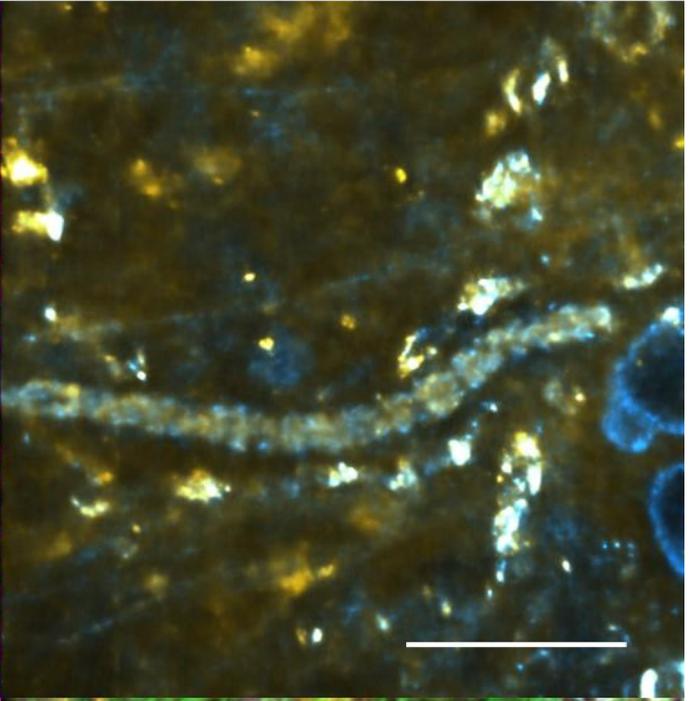
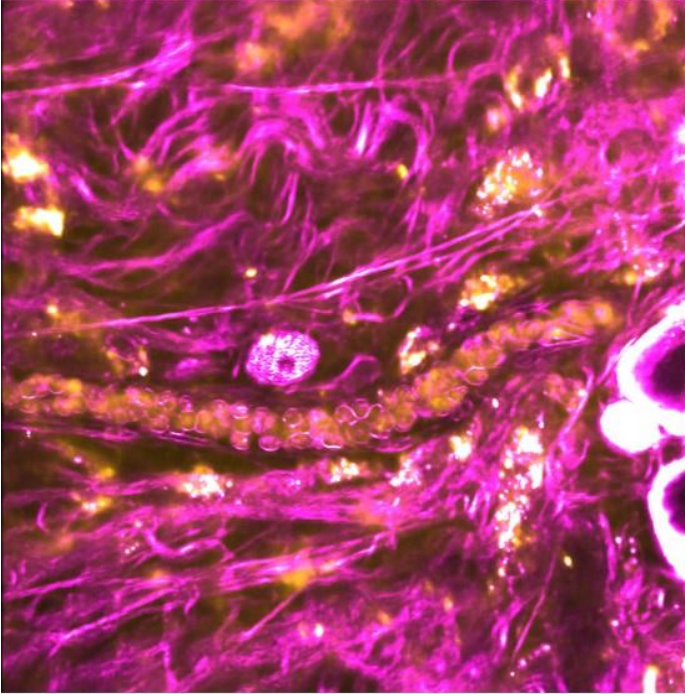 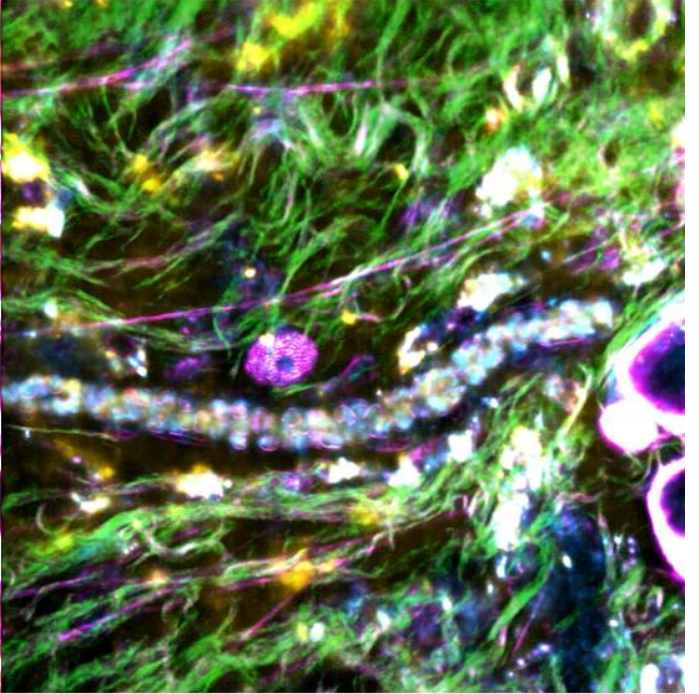

Fig. S3. intravital eSLAM imaging of mouse skin flap at one instance showing the presence of intracellular 2PAF/yellow, 3PAF/cyan, and THG/magenta signals in different parts of single biconcave disk-shaped blood cells. Scale bar: 50 μm.

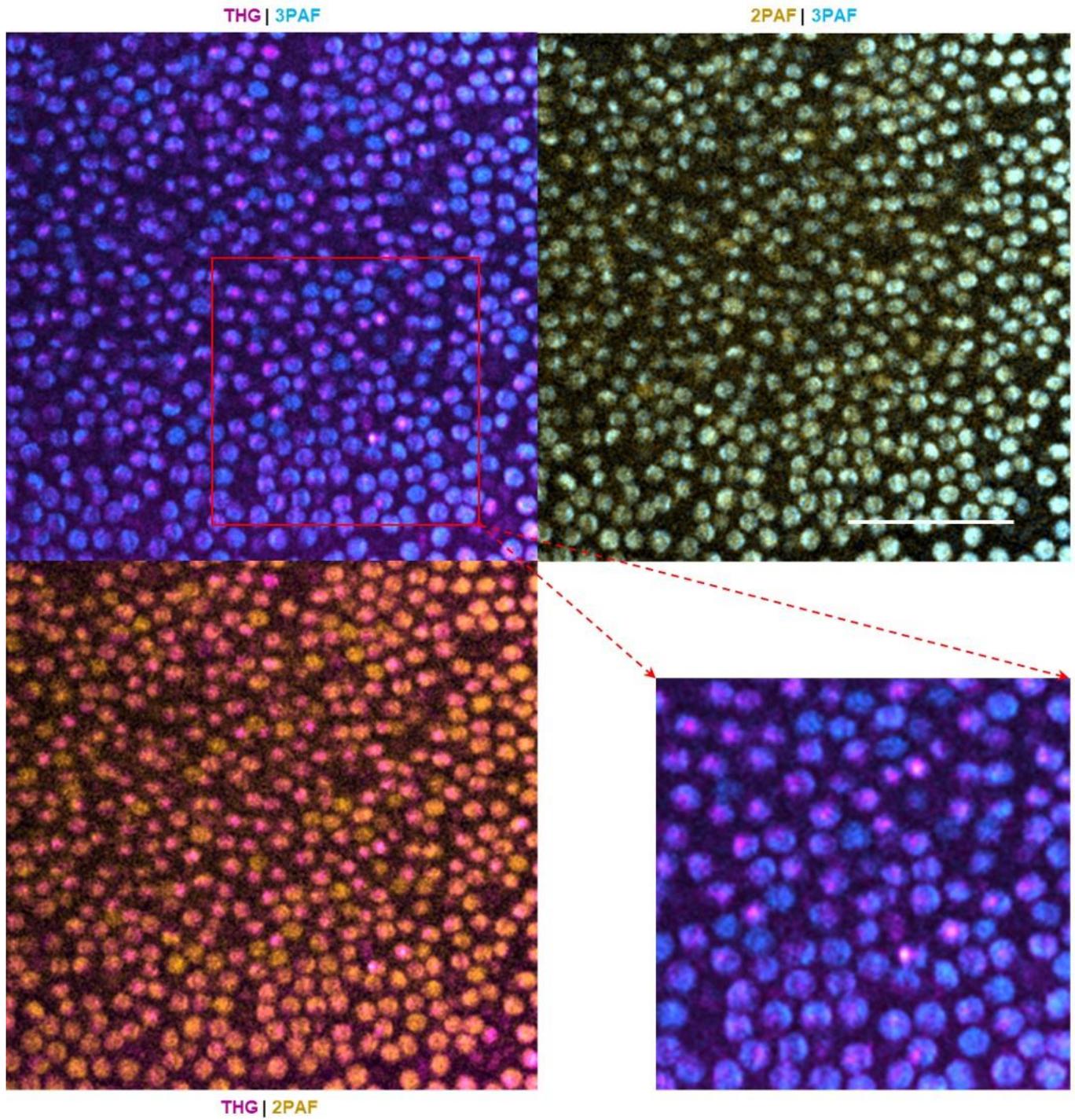

Fig. S4. Mouse blood smear from eSLAM imaging confirms the presence of intracellular 2PAF/yellow, 3PAF/cyan, and THG/magenta signals in different parts of single blood cells. Scale bar: 50 μm.

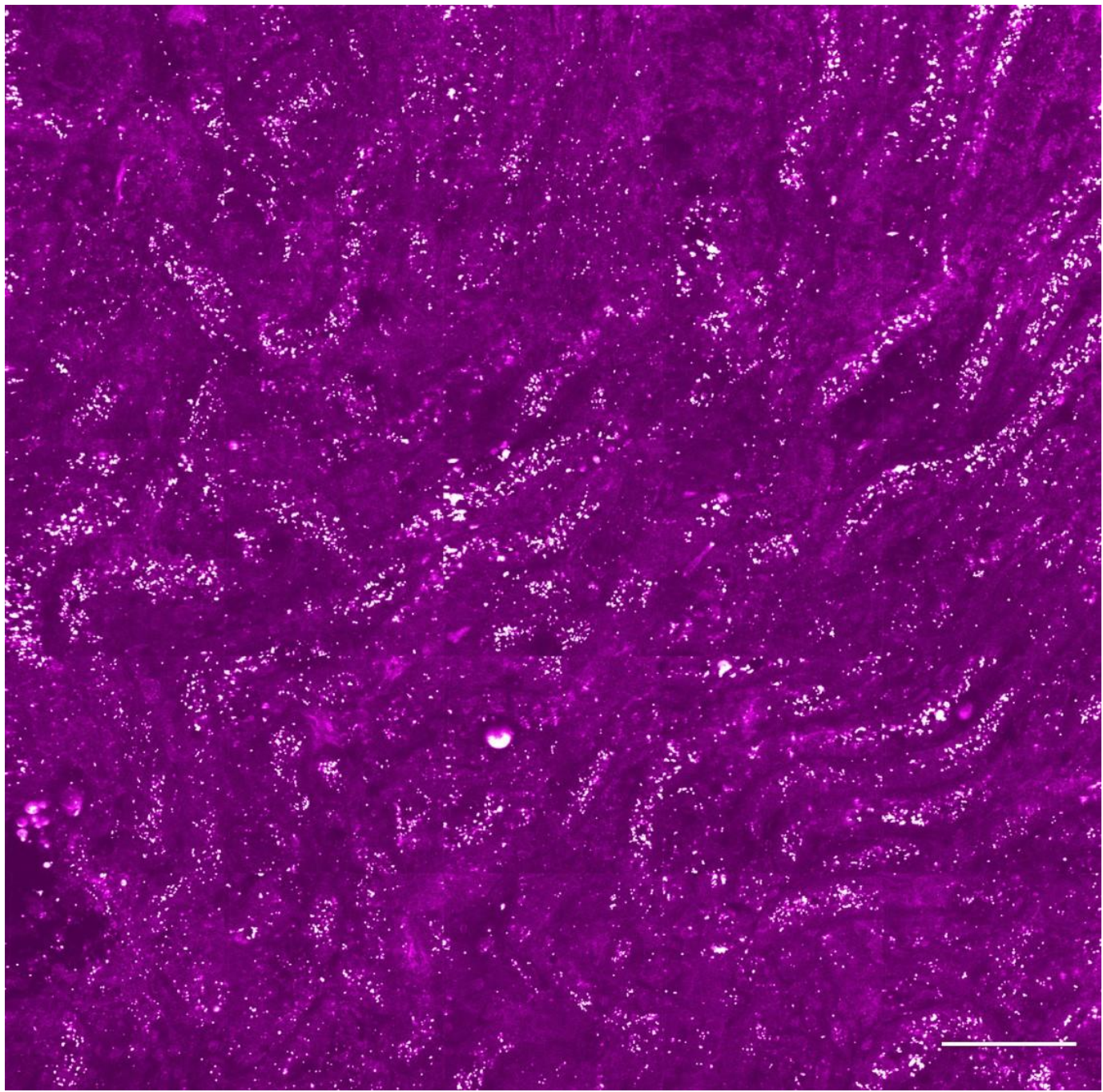

Fig. S5a. THG image of ex vivo mouse kidney from mosaic eSLAM imaging. Scale bar: 100 μm.

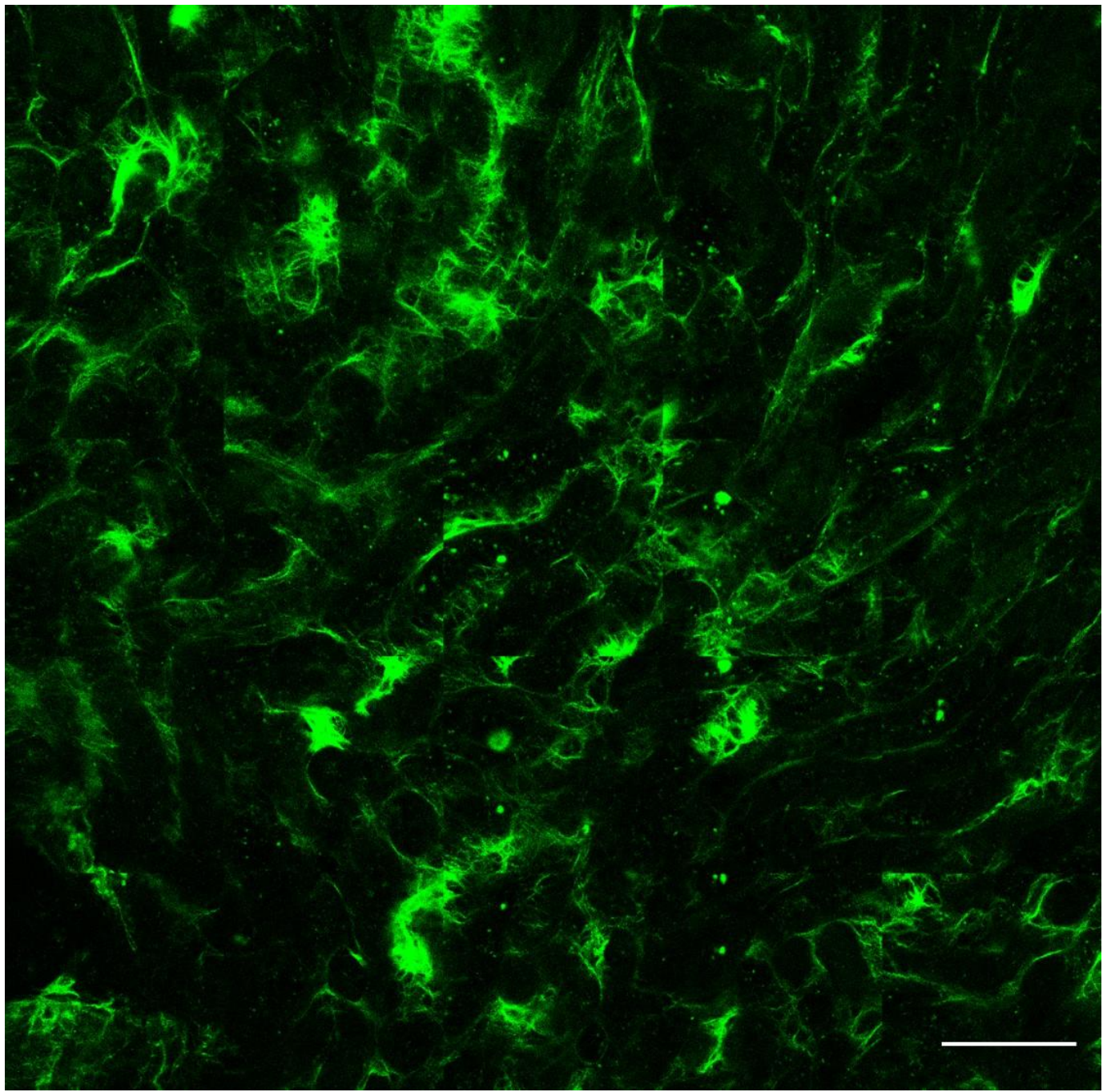

Fig. S5b. SHG image of ex vivo mouse kidney from mosaic eSLAM imaging. Scale bar: 100 μm.

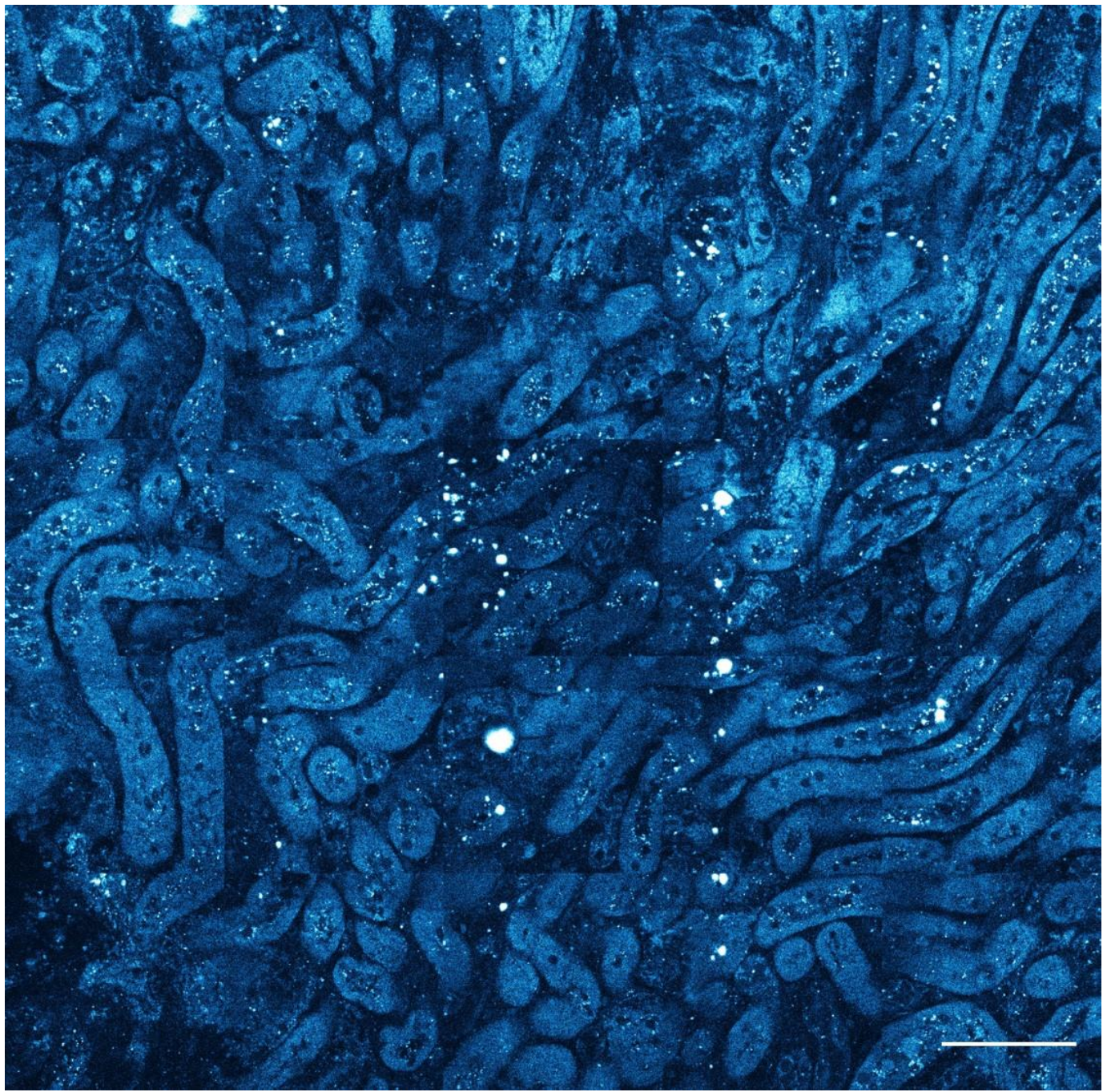

Fig. S5c. 3PAF intensity image of ex vivo mouse kidney from mosaic eSLAM imaging. Scale bar: 100 μm.

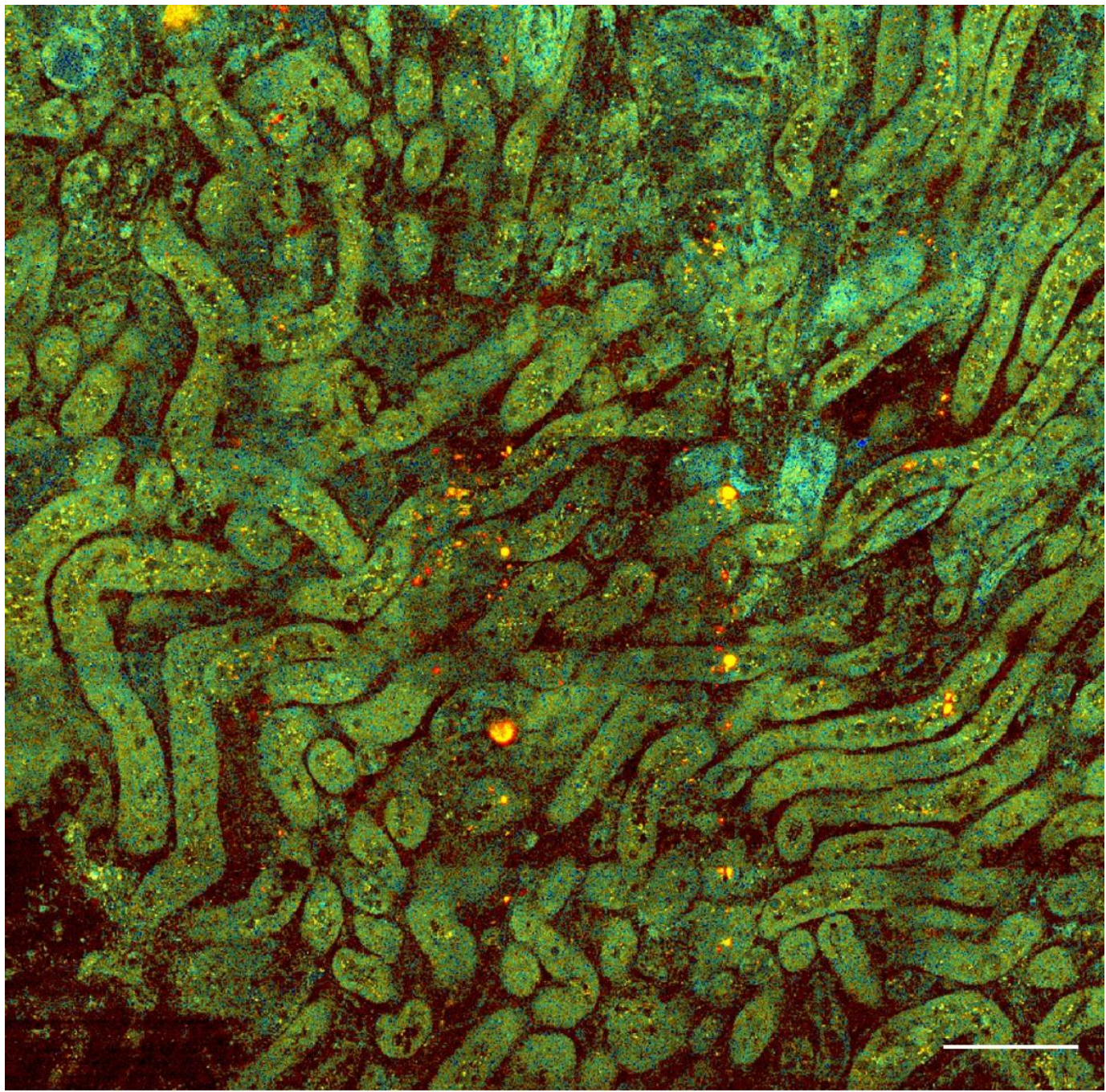
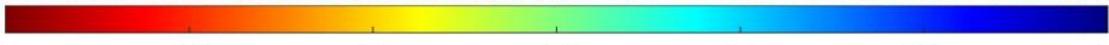

Fig. S5d. 3PAF lifetime image of ex vivo mouse kidney from mosaic eSLAM imaging. Scale bar: 100 μm.

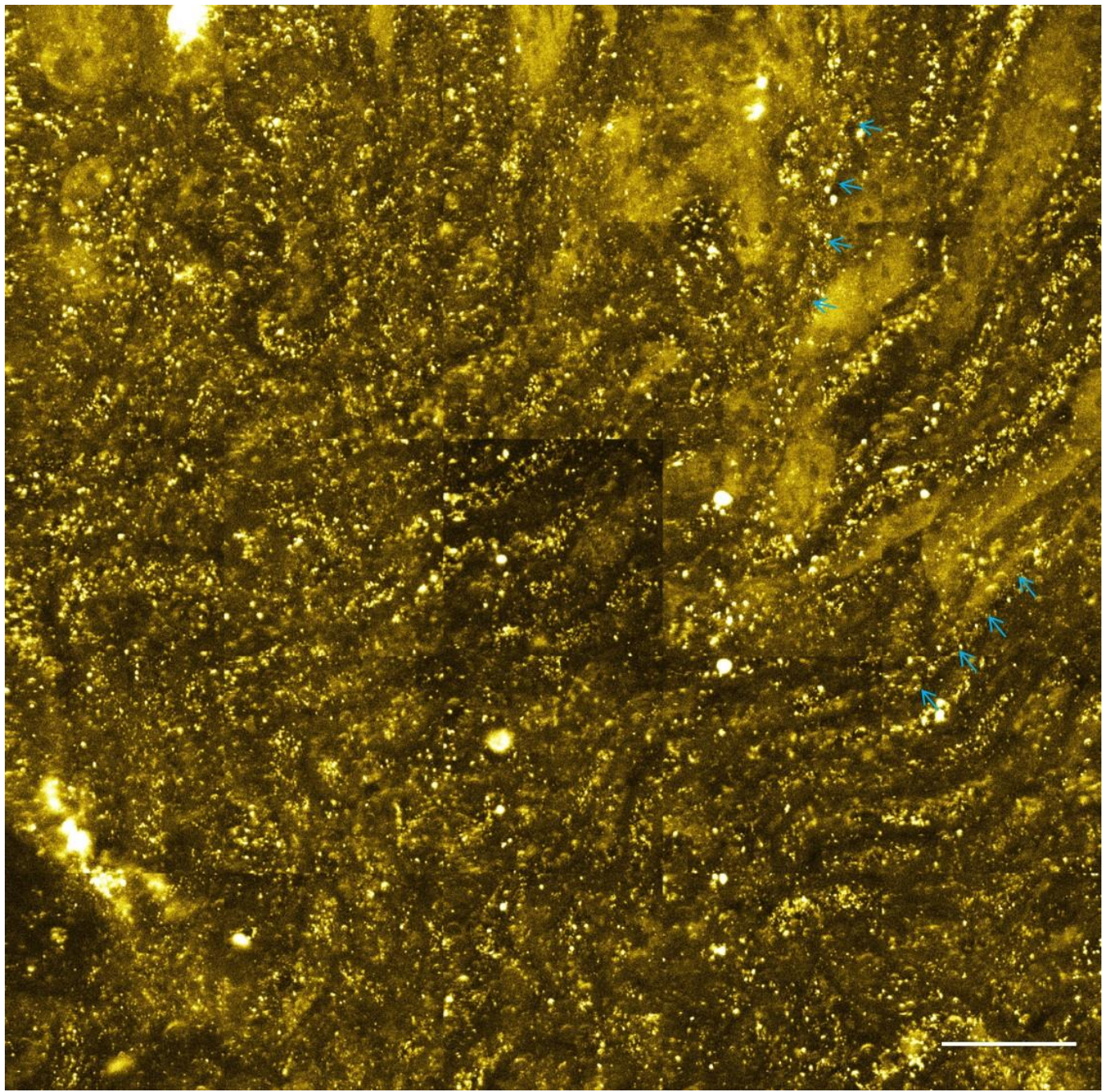

Fig. S5e. 2PAF intensity image of ex vivo mouse kidney from mosaic eSLAM imaging. Scale bar: 100 μm.

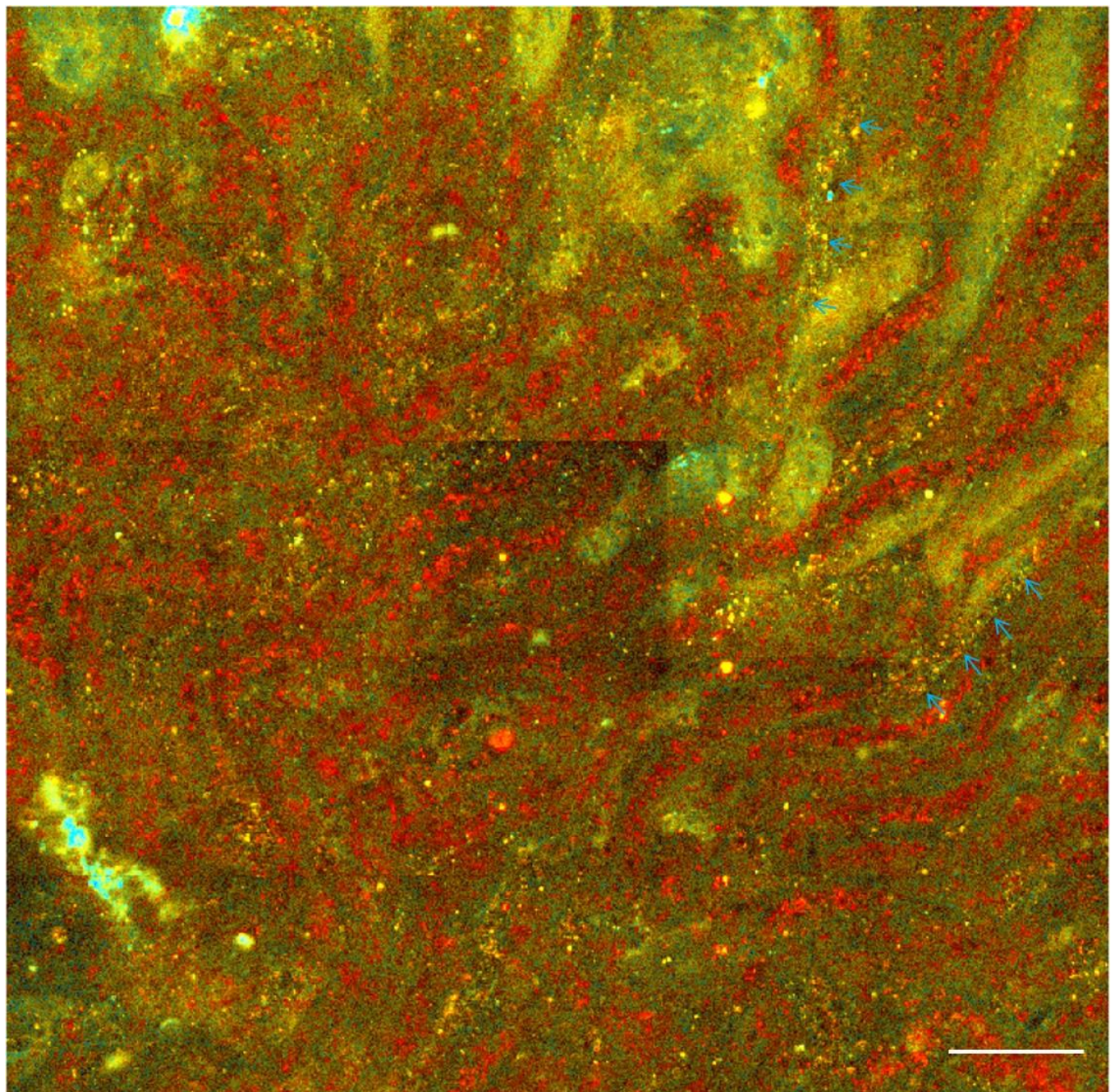

Fig. S5f. 2PAF lifetime image of ex vivo mouse kidney from mosaic eSLAM imaging. Scale bar: 100 μm.